\documentclass[aps,twocolumn,prl,superscriptaddress,amsmath,amssymb]{revtex4-1}
\newcommand{\means}[1]{\langle#1\rangle}

\ifx\pdfoutput\undefined
\usepackage[dvipdfmx]{graphicx}
\usepackage[dvipdfmx]{hyperref}
\else
\usepackage{graphicx}
\usepackage{hyperref}
\fi

\usepackage{txfonts}
\usepackage[T1]{fontenc}
\usepackage{xspace}

\usepackage{dcolumn}
\usepackage{bm}
\usepackage{amsmath}
\usepackage{booktabs}
\usepackage[usenames]{color}
\usepackage{xcolor}

\usepackage{ulem}

\begin{document}
\let\emph\textit

%\draft
%\preprint{?????}
\title{
Thermal Transport in the Kitaev Model
}

\author{Joji Nasu}
\affiliation{Department of Physics, Tokyo Institute of Technology, Meguro, Tokyo 152-8551, Japan}
\author{Junki Yoshitake}
\affiliation{Department of Applied Physics, University of Tokyo, Bunkyo, Tokyo 113-8656, Japan}
\author{Yukitoshi Motome}
\affiliation{Department of Applied Physics, University of Tokyo, Bunkyo, Tokyo 113-8656, Japan}

 \date{\today}
 \begin{abstract}
  In conventional insulating magnets, heat is carried by magnons and phonons.
  In contrast, when the magnets harbor a quantum spin liquid state, emergent quasiparticles from the fractionalization of quantum spins can carry heat.
  Here, we investigate unconventional thermal transport yielded by such exotic carriers, in both longitudinal and transverse components, for the Kitaev model, whose ground state is exactly shown to be a quantum spin liquid with fractional excitations described as itinerant Majorana fermions and localized $Z_2$ fluxes.
  We find that the longitudinal thermal conductivity exhibits a broad peak at very different temperatures between the zero and nonzero frequency components, reflecting the spin fractionalization.
  On the other hand, the transverse thermal conductivity induced by the magnetic field shows nonmonotonic temperature dependence, due to thermal excitations of the localized $Z_2$ fluxes.
  In the low-temperature limit, the temperature-linear coefficient rapidly approaches a quantized value, as expected from the topologically nontrivial nature of itinerant Majorana fermions.
  The characteristic behaviors provide experimentally-accessible evidences of fractional excitations in the proximity to Kitaev quantum spin liquid.
 \end{abstract}

%\pacs{75.25.Dk,75.70.Tj,75.10.Jm,75.30.Et}
%\pacs{75.10.Kt,75.10.Jm,75.30.Et}
%75.10.Kt	Quantum spin liquids, valence bond phases and related phenomena
%75.10.Jm	Quantized spin models, including quantum spin frustration
%03.67.Pp	Quantum error correction and other methods for protection against decoherence
%71.10.Pm	Fermions in reduced dimensions (anyons, composite fermions, Luttinger liquid, etc.)
%03.67.Lx	Quantum computation architectures and implementations
%75.30.Et	Exchange and superexchange interactions
%71.70.Ej	Spin-orbit coupling, Zeeman and Stark splitting, Jahn-Teller effect
%71.35.Lk	Collective effects (Bose effects, phase space filling, and excitonic phase transitions)
%73.43.Nq	Quantum phase transitions (see also 64.70.Tg Quantum phase transitions in equations of state, phase equilibria and phase transitions)
%75.30.Kz	Magnetic phase boundaries (including classical and quantum magnetic transitions, metamagnetism, etc.)
%74.25.Bt	Thermodynamic properties

%75.25.Dk 	Orbital, charge, and other orders, including coupling of these orders
%75.30.Et 	Exchange and superexchange interactions (see also 71.70.Gm Exchange interactions)
%75.47.Lx 	Magnetic oxides
%75.70.Tj	Spin-orbit effects (see also 71.70.Ej Spin-orbit coupling, Zeeman and Stark splitting, Jahn-Teller effect)

\maketitle

  Insulating magnets provide a paradigmatic playground for quantum many-body effects in the spin degree of freedom of electrons in solids.
  In conventional magnets, the elementary excitation is known as magnons, describing the collective modes associated with long-range magnetic orders.
  Once the ordering is suppressed by strong quantum fluctuations, however, the magnets may possess exotic excitations yielded by the quantum many-body effects.
  Fractional quasiparticles in quantum spin liquids (QSLs), where any spontaneous symmetry breaking does not appear even at zero temperature ($T$), are archetypal examples of such exotic excitations~\cite{Wen1991,Lacroix2011C16}.
  For instance, in the celebrated resonating-valence-bond state, the low-energy excitations are described by spinons, spin-$1/2$ fermionic quasiparticles, which may bring about an emergent Fermi surface despite insulating magnets~\cite{Anderson1973153,fazekas1974ground}.
  Considerable efforts have been devoted to experimental observation of such itinerant nature, e.g., in the low-$T$ behavior of the specific heat and the thermal transport~\cite{ISI:000278318600025,PhysRevLett.112.177201,watanabe2016emergence}.
  Nevertheless, identifying the fractionalization remains elusive, because of not only obstacles in experiments but also the lack of detailed theoretical information.

  Recently, a quantum spin model, whose ground state is exactly shown to be a QSL, was proposed by Kitaev~\cite{Kitaev2006}.
  The elementary excitations in the Kitaev QSL are described by two types of quasiparticles emergent from the fractionalization of quantum spins: itinerant Majorana fermions and localized $Z_2$ fluxes.
  Moreover, it was pointed out that this Kitaev model may give a good description of spin-orbital entangled magnets~\cite{PhysRevLett.102.017205}, such as $A_2$IrO$_3$ ($A=$ Li, Na) and $\alpha$-RuCl$_3$.
  On this basis, precursors of the Kitaev QSL have been investigated in these materials, e.g., by the neutron and Raman scattering measurements~\cite{banerjee2016proximate,PhysRevLett.114.147201}, in comparison with the theoretical calculations~\cite{PhysRevLett.112.207203,PhysRevLett.113.187201,PhysRevB.93.174425,Nasu2016nphys,Song2016,yoshitake2016}.
  Very recently, some attempts have been made to grasp the itinerant nature of the Majorana fermions by measuring the thermal conductivity~\cite{hirobe2016pre,leahy2016pre,Hentrich2017pre}.
  While theoretical works have been done for the thermal transport in one-dimensional Kitaev systems~\cite{PhysRevB.93.214425,Metavitsiadis2016pre} and that carried by magnons in magnetically ordered states in the Kitaev-Heisenberg model~\cite{Stamokostas2017}, the thermal conductivity owing to the Majorana fermions emergent in the two-dimensional Kitaev QSL remains unclear.
  Furthermore, an applied magnetic field can change the topology of the Majorana fermion states~\cite{Kitaev2006}, which may lead to the quantized transverse conductivity in the low-$T$ limit.
  Thus, it is highly desired to provide theoretical inputs on the thermal transport in both longitudinal and transverse components for the Kitaev model as a canonical reference.

  In this Letter, we investigate the thermal transport originating from the emergent fractional quasiparticles in the Kitaev model.
  Using quantum Monte Carlo (QMC) simulations, we calculate the $T$ dependences of both longitudinal and transverse thermal conductivities, with and without the weak magnetic field.
  We find that, in the absence of the magnetic field, the longitudinal zero(nonzero)-frequency component exhibits a broad peak around the higher(lower)-$T$ peak of the specific heat.
  The longitudinal responses are suppressed by an applied magnetic field.
  On the other hand, the transverse component is induced by the magnetic field, and that divided by $T$ shows a nonmonotonic $T$ dependence with rapidly approaching a quantized value at low $T$.
  We discuss these peculiar behaviors associated with thermal excitations of the fractional quasiparticles.

\begin{figure}[t]
 \begin{center}
  \includegraphics[width=\columnwidth,clip]{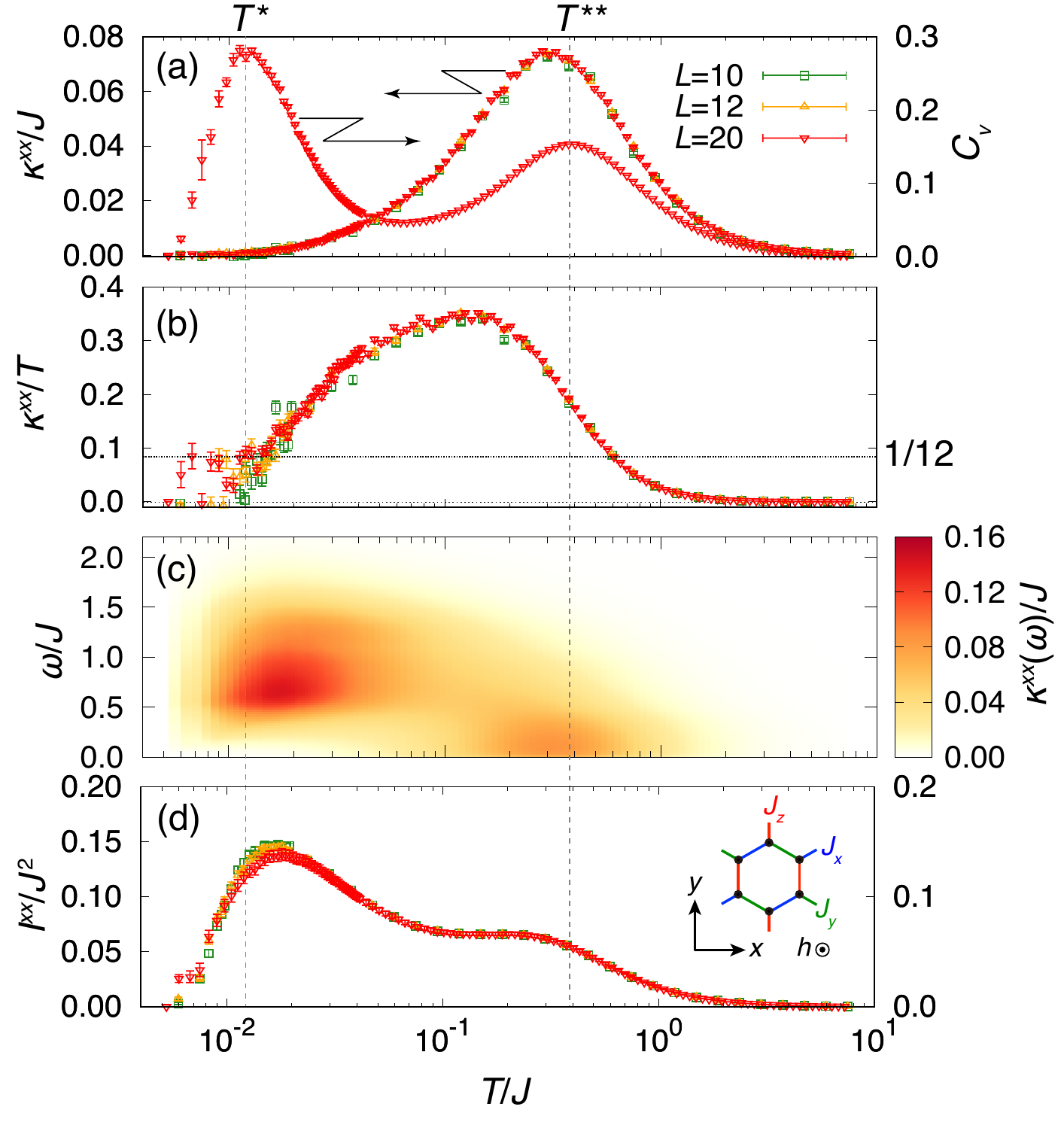}
  \caption{
  (a) Longitudinal thermal conductivity, $\kappa^{xx}=\lim_{\omega\to 0} \kappa^{xx}(\omega)$, plotted with the specific heat $C_v$ and (b) $\kappa^{xx}/T$ as functions of $T$.
  $T^*$ and $T^{**}$ are two crossover temperatures, determined from the broad peaks in $C_v$.
  (c) Contour map of $\kappa^{xx}(\omega)$ on the $T$-$\omega$ plane calculated for the $L=20$ cluster.
  (d) Integrated intensity $I^{xx}=\int_0^\infty \kappa^{xx}(\omega)d\omega$.
  The inset of (d) represents the honeycomb lattice on the $xy$ plane where the Kitaev model is defined in an applied magnetic field $h$ along the $z$ direction [Eq.~(\ref{eq:H})].
  The different bond colors illustrate three different types of bonds in the Kitaev model.
}
  \label{kl}
 \end{center}
\end{figure}

  We consider the Kitaev model in an external magnetic field applied perpendicular to the honeycomb plane.
  The spin axis is taken by following Ref.~\cite{PhysRevLett.102.017205}, so as to be relevant to candidate materials like $A_2$IrO$_3$ ($A=$ Li, Na) and $\alpha$-RuCl$_3$.
  The Hamiltonian is given by
\begin{eqnarray}
 {\cal H} =-J\sum_{\gamma=x,y,z}\sum_{\means{jj'}_\gamma}S_j^\gamma S_{j'}^\gamma -h\sum_j(S_j^x+S_j^y+S_j^z),
 \label{eq:H}
\end{eqnarray}
where $\means{jj'}_\gamma$ represents a nearest-neighbor pair on one of three sets of inequivalent bonds, $\bm{S}_j=(S_j^x,S_j^y,S_j^z)$ is an $S=1/2$ operator at position $\bm{r}_j$, $J$ is the exchange constant assumed to be isotropic for three types of bonds, and $h$ represents the magnetic field strength; see the inset of Fig.~\ref{kl}(d).
  In the absence of the magnetic field ($h=0$), the ground state of the Kitaev model is exactly obtained by introducing itinerant Majorana fermions and localized $Z_2$ fluxes $W_p$, the latter of which are defined for each hexagonal plaquette $p$ on the honeycomb lattice~\cite{Kitaev2006}.
  The ground state is given by all $W_p=+1$ (flux-free state), and the system is in a gapless QSL phase where the itinerant Majorana fermion spectrum forms the massless Dirac nodes.
  In contrast, there is a nonzero gap $\Delta\sim 0.065J$ in the excitation of the localized $Z_2$ fluxes.

  When $h$ is small enough compared to $\Delta$, one can derive an effective model by using the third-order perturbation, whose Hamiltonian is given by~\cite{Kitaev2006}
\begin{eqnarray}
 \tilde{{\cal H}}=-J \sum_{\gamma=x,y,z}\sum_{\means{jj'}_\gamma}S_j^\gamma S_{j'}^\gamma -\tilde{h}\sum_{[jj''j']_{\alpha\beta\gamma}}S_j^\alpha S_{j''}^\beta S_{j'}^\gamma,\label{eq:1}
\end{eqnarray}
where the effective magnetic field $\tilde{h}\sim h^3/\Delta^2$;
  $[jj''j']_{\alpha\beta\gamma}$ represents neighboring three sites, where the neighboring pair $jj''$ ($j''j'$) are located on an $\alpha$ ($\gamma$) bond and $\beta$ is taken to be neither $\alpha$ nor $\gamma$.
  The Hamiltonian $\tilde{{\cal H}}$ is exactly soluble for all $\tilde{h}$, while ${\cal H}$ in Eq.~(\ref{eq:H}) is not for $h\neq 0$~\cite{Kitaev2006}.
  This is shown by, e.g., introducing two kinds of Majorana fermions $c_j$ and $\bar{c}_j$~\cite{PhysRevB.76.193101,PhysRevLett.98.087204,1751-8121-41-7-075001}, which enable to rewrite the Hamiltonian into a bilinear form in terms of $c_i$ as
$\tilde{\cal H}=\frac{1}{2}\sum_{jj'}c_{j}A_{jj'}(\{\eta_b\}) c_{j}$;
  $A(\{\eta_b\})$ is a pure-imaginary Hermite matrix dependent on $\eta_b=i\bar{c}_j\bar{c}_{j'}$, which is a $Z_2$ conserved quantity taking $\pm 1$ on the $z$ bond $b=\means{jj'}_z$~\footnote{See Supplemental Material.}.
  The flux in the plaquette $p$ is given by $W_p=\eta_{b_1}\eta_{b_2}$, where $b_1$ and $b_2$ are the $z$ bonds included in the hexagon $p$.
  The three-spin term in Eq.~(\ref{eq:1}) turns into second-neighbor hopping of $c_j$ in the bilinear Hamiltonian.
  Interestingly, this hopping term opens a gap in the Dirac spectrum of the Majorana fermion system and yields a chiral edge mode within the gap~\cite{Kitaev2006}, similar to the Haldane model showing the quantum anomalous Hall effect in a zero magnetic field~\cite{Haldane1988}.
  This topological nature was confirmed for the original model in Eq.~(\ref{eq:H})~\cite{Jiang2011}.
  We note that, in addition to the second term in Eq.~(\ref{eq:1}), the third-order perturbation in terms of $h$ leads to another three-spin interactions described by interactions between the Majorana fermions $c$, which is supposed be irrelevant to the Dirac gap opening and omitted in the following analysis~\cite{Kitaev2006,Hermanns2015}.

  The bilinear Majorana representation for Eq.~(\ref{eq:1}) admits us to perform QMC simulations without the negative sign problem~\cite{PhysRevLett.113.197205,PhysRevB.92.115122,Nasu2016nphys}.
  To evaluate the thermal conductivity, we introduce the energy polarization operator defined as $\bm{P}_E=\sum_{jj'}\frac{\bm{r}_{j}+\bm{r}_{j'}}{2}\tilde{\cal H}_{jj'}$, where $\tilde{{\cal H}}_{jj'}=\frac{1}{2}c_{j}A_{jj'} c_{j'}$~\cite{PhysRevLett.104.066403}.
  We set the Boltzmann constant $k_B$ and the reduced Planck constant $\hbar$ to be unity.
  The energy current is introduced as
  $\bm{J}_E=\frac{\partial \bm{P}_E}{\partial t}=i[\tilde{{\cal H}},\bm{P}_E]$.
  In the Majorana fermion system, as the chemical potential is always fixed to zero regardless of the configuration of $\{\eta_b\}$, the energy current is equivalent to the heat current $\bm{J}_Q$.
  The thermal conductivity $\kappa^{\mu\nu}$ ($\mu,\nu=x,y$) is obtained by using the Kubo formula as $\kappa_{\rm Kubo}^{\mu\nu}(\omega)=\frac{1}{TV}\int_0^\infty dt e^{i(\omega+i\delta) t}\int_0^\beta d\lambda\means{J_Q^\mu(-i\lambda)J_Q^\nu(t)}$, where $J_Q^\mu(t)=e^{i\tilde{\cal H}t}J_Q^\mu e^{-i\tilde{\cal H}t}$, $\beta=1/T$ is the inverse temperature, $V$ is the volume of the system, and $\delta$ is a positive infinitesimal constant.
  While the longitudinal component is simply given by $\kappa^{\mu\mu}(\omega)=\kappa_{\rm Kubo}^{\mu\mu}(\omega)$, the transverse component $\kappa^{\mu\nu}(\omega)$ needs a contribution from ``the gravitational magnetization'' in addition to $\kappa_{\rm Kubo}^{\mu\nu}(\omega)$~\cite{Nomura2012,doi:10.7566/JPSJ.82.023602}.
  We calculate $\kappa^{\mu\nu}(\omega)$ for about 300 samples taken from the 40,000 MC steps after 10,000 MC steps for thermalization, in the $30\times 30$ supercell of the $2L^2$ cluster used in the MC simulations.
  The details of the calculation are given in Supplemental Material~\footnotemark[1].

  First, we examine the longitudinal component of the thermal conductivity $\kappa^{xx}$($=\kappa^{yy}$) in the absence of magnetic field $\tilde{h}=0$.
  Figure~\ref{kl}(a) shows the $T$ dependence of $\kappa^{xx} = \lim_{\omega\to 0} \kappa^{xx}(\omega)$~\footnotemark[1].
  We also display the specific heat $C_v$ in Fig.~\ref{kl}(a).
  $C_v$ has two broad peaks at $T^*\simeq 0.012J$ and $T^{**}\simeq 0.375J$ due to thermal fractionalization~\cite{PhysRevB.92.115122}: the low-$T$ crossover at $T^*$ comes from the release of a half of $\ln 2$ entropy related to the localized $Z_2$ fluxes $W_p$, while the high-$T$ one at $T^{**}$ is by the rest half from the itinerant Majorana fermions.
  In contrast, we find that $\kappa^{xx}$ exhibits only a single broad peak near $T^{**}$.
  The result is far from the conventional wisdom that predicts $\kappa^{xx} \propto C_v$.
  This discrepancy is a direct consequence of the thermal fractionalization of quantum spins.
  Among the fractional quasiparticles, only the itinerant Majorana fermions can carry heat.
  Hence, $\kappa^{xx}$ has a substantial contribution only near $T^{**}$ where the itinerant Majorana fermions release their entropy.

  We also present $\kappa^{xx}/T$ in Fig.~\ref{kl}(b).
  As decreasing $T$, this quantity increases from zero around $T^{**}$, and decreases with approaching $T^*$ after showing a broad hump at $T\sim 0.1J$.
  Below $T^{*}$, $\kappa^{xx}/T$ appears to approach $\sim 1/12$, although the size dependence and statistical errors become comparatively large.
  The asymptotic behavior might be related to the minimum conductivity in graphene with disorder~\cite{Ludwig1994,Ziegler2007}, as thermally excited $Z_2$ fluxes are regarded as scatterers for the itinerant Majorana fermions moving on the honeycomb lattice~\footnote{The Wiedemann-Franz law is not violated in the present case with noninteracting Majorana fermions coupled with thermally fluctuating $Z_2$ fluxes, in contrast to interacting Dirac electron systems~\cite{Principi2015,Lucas2016}. The presence of the minimum $T$-linear coefficient of the thermal conductivity was also discussed in the Majorana fermion system with long-range disorder~\cite{nakai2014}.}.
  We note that the thermal conductivity by Majorana fermions is halved from that by electrons~\cite{Nomura2012}.

  Figure~\ref{kl}(c) shows the contour map of $\kappa^{xx}(\omega)$ on the $T$-$\omega$ plane.
  $\kappa^{xx}(\omega)$ has a low-$\omega$ weight around $T^{**}$, which gives the peak of the $\omega \to 0$ limit, i.e., $\kappa^{xx}$, shown in Fig.~\ref{kl}(a).
  On the other hand, the weight is shifted to a higher $\omega$ region as lowering $T$, and enhanced to form a broad peak at $\omega\sim J$ around $T^*$.
  The weight is reduced rapidly for further lowering $T$ below $T^*$; we note that $\kappa^{xx}(\omega)$ completely vanishes at $T=0$ as the Kitaev Hamiltonian commutes with the energy current in the flux-free ground state~\footnotemark[1].
  We plot the integrated thermal conductivity defined by $I^{xx}=\int_0^{\infty}\kappa^{xx}(\omega)d\omega$ in Fig.~\ref{kl}(d).
  As increasing $T$ from $T=0$, $I^{xx}$ rapidly grows from zero and forms a broad peak around $T^*$ originating from the contribution at $\omega\sim J$ of $\kappa^{xx}(\omega)$, while it decreases in the higher-$T$ region after showing a shoulder around $T^{**}$.
  Thus, the nonzero-frequency thermal response is a good measure for the $Z_2$ flux excitations.

\begin{figure}[t]
 \begin{center}
  \includegraphics[width=\columnwidth,clip]{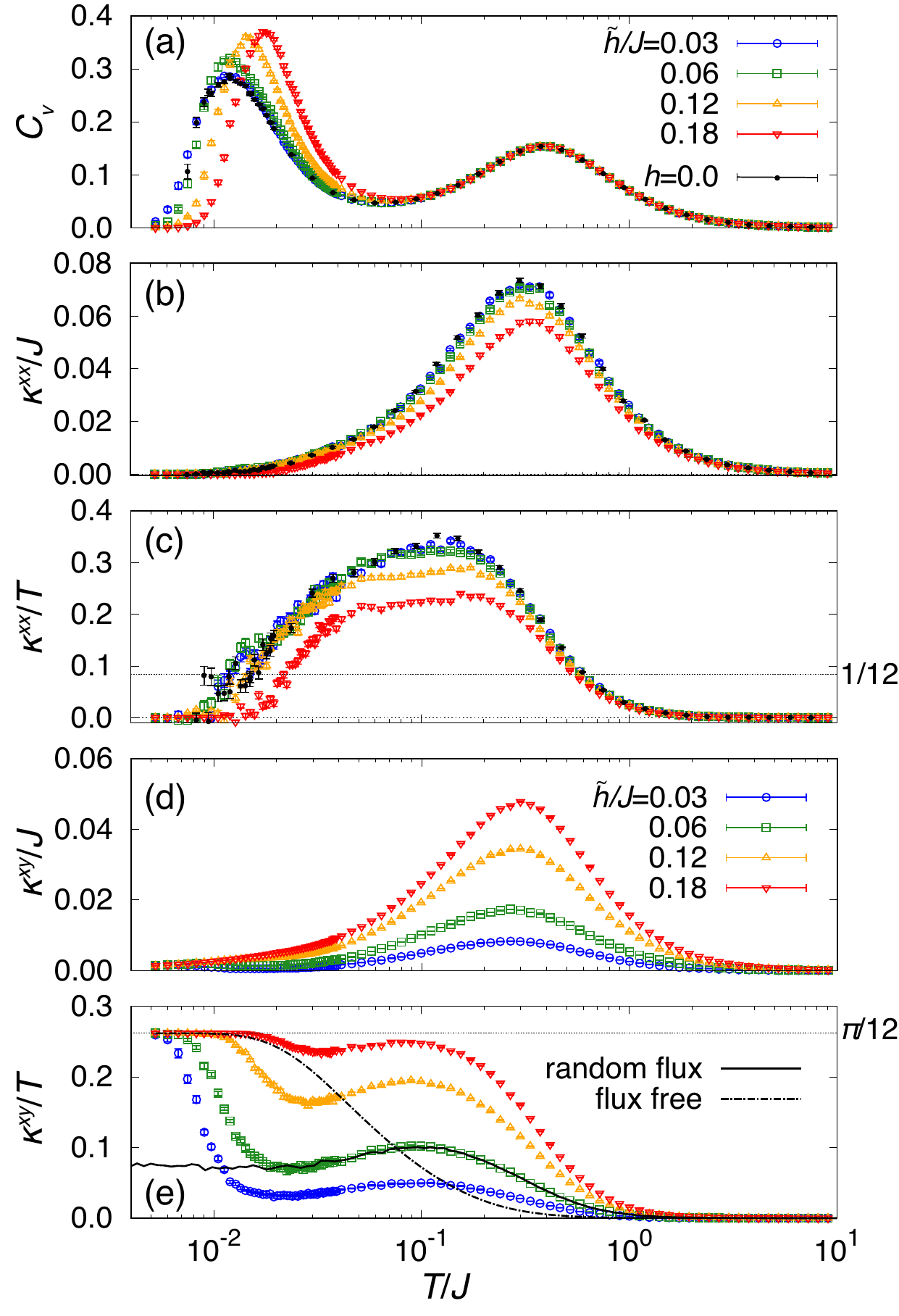}
  \caption{
  (a) $C_v$, (b) $\kappa^{xx}$, (c) $\kappa^{xx}/T$, (d) $\kappa^{xy}$, and (e) $\kappa^{xy}/T$ in the magnetic field $\tilde{h}$.
  The data are calculated for $L=8$, $10$, and $12$, but we show only the data for $L=12$ as they are indistinguishable within the statistical errors.
  In (e), we also plot $\kappa^{xy}/T$ calculated for the flux-free (dashed-dotted) and random flux (solid) cases at $\tilde{h}/J=0.06$.
}
  \label{hall}
 \end{center}
\end{figure}

  Next, we show the results in the presence of the magnetic field $\tilde{h}$ in Eq.~(\ref{eq:1}).
  Figure~\ref{hall}(a) presents the specific heat.
  The high-$T$ peak at $T^{**}$ hardly changes by the magnetic field, while the low-$T$ peak at $T^*$ slowly shifts to the high-$T$ side.
  Figure~\ref{hall}(b) shows that the longitudinal thermal conductivity $\kappa^{xx}$ decreases by the magnetic field with retaining the single peak structure, whereas $C_v$ is almost unchanged in the high-$T$ region.
  The effect of the magnetic field is also seen in $\kappa^{xx}/T$, as shown in Fig.~\ref{hall}(c).
  In addition to the suppression due to $\tilde{h}$, $\kappa^{xx}/T$ at $T\to 0$ vanishes, reflecting the Dirac gap in the Majorana spectrum.

  As mentioned above, in the gapped state for nonzero $\tilde{h}$, the system has a chiral edge mode inside the gap, which may lead to nontrivial topological phenomena such as off-diagonal transport~\cite{Kitaev2006}, as in the massive Dirac fermion systems~\cite{nomura2011}.
  We here calculate the transverse thermal conductivity $\kappa^{xy}$ [Fig.~\ref{hall}(d)].
  Note that $\kappa^{xy}$ is directly obtained without any extrapolation in terms of $\omega$, in contrast to $\kappa^{xx}$~\footnotemark[1].
  $\kappa^{xy}$ shows a broad peak around $T^{**}$ similar to $\kappa^{xx}$.
  However, the $\tilde{h}$ dependence of $\kappa^{xy}$ is distinct from that of $\kappa^{xx}$; the peak is enhanced by the magnetic field continuously from $\tilde{h}=0$.

  Figure~\ref{hall}(e) plots $\kappa^{xy}/T$.
  The $T$ dependence is nonmonotonic; while decreasing $T$, $\kappa^{xy}/T$ increases from zero around $T^{**}$, and shows a hump at $T\sim 0.1J$ similar to $\kappa^{xx}/T$ in Fig.~\ref{hall}(c), and eventually, it arises again and converges a quantized value $\pi/12$ as $T\to 0$.
  This peculiar $T$ dependence is understood by thermal excitations of the $Z_2$ fluxes.
  In Fig.~\ref{hall}(e), we present $\kappa^{xy}/T$ for the flux-free state (all $W_p=+1$) and the random $W_p$ configuration at $\tilde{h}=0.06J$.
  For the latter, we evaluate $\kappa^{xy}/T$ from $10^3$ random configurations of $\{\eta_b\}$.
  When we assume the flux-free configuration, $\kappa^{xy}/T$ monotonically decreases from the quantized value with increasing $T$.
  The QMC result deviates from this behavior with more rapid decrease around $T^{*}$ where the $Z_2$ fluxes are thermally excited, as shown in Fig.~\ref{hall}(e).
  On the other hand, for the random configuration, $\kappa^{xy}/T$ shows a hump around $T/J\sim 0.1$, which well accounts for the QMC data.
  This analysis indicates that the nonmonotonic $T$ dependence of $\kappa^{xy}/T$ is yielded by thermal excitation of $Z_2$ fluxes.

\begin{figure}[t]
 \begin{center}
  \includegraphics[width=\columnwidth,clip]{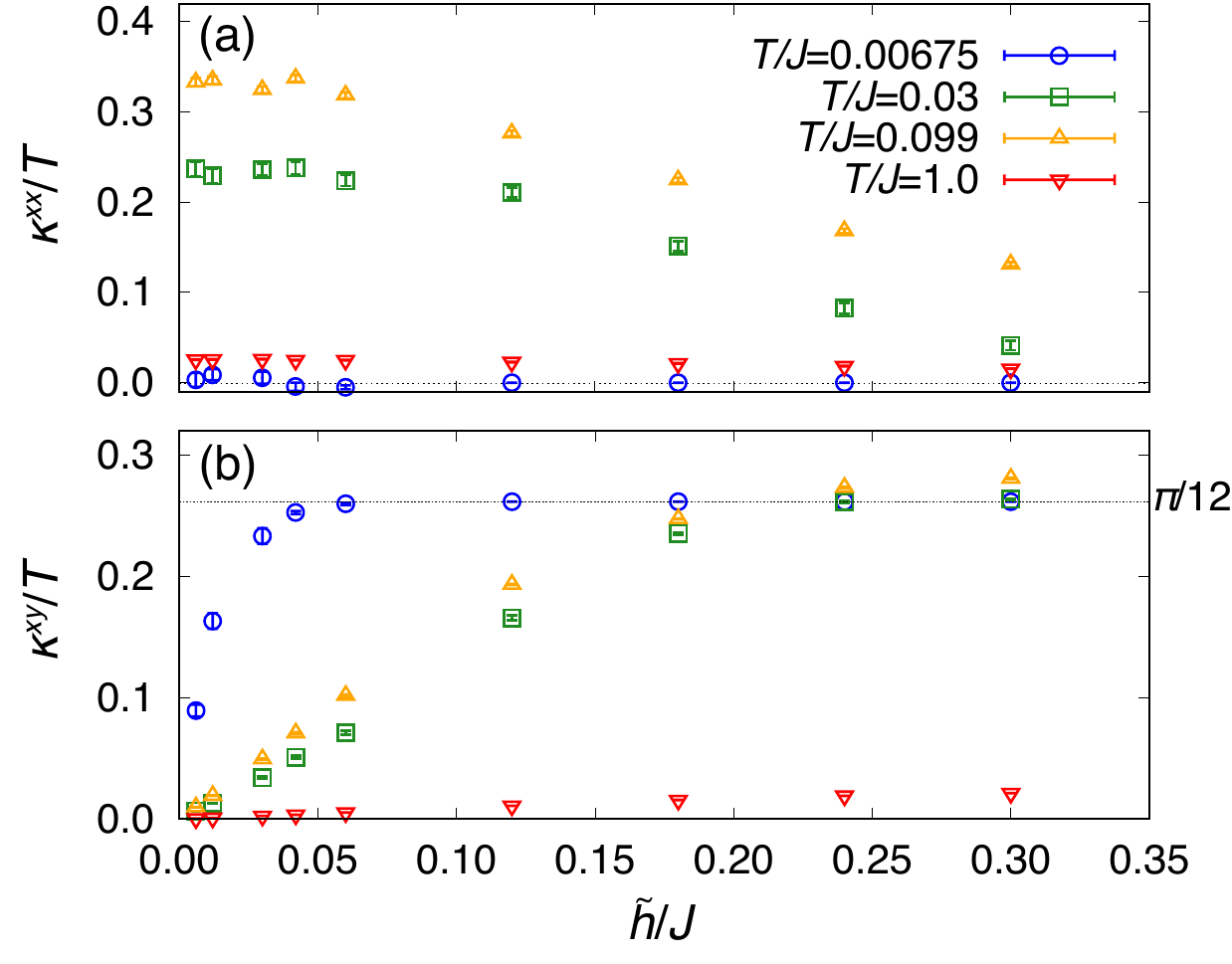}
  \caption{
$\tilde{h}$ dependences of (a) $\kappa^{xx}/T$ and (b) $\kappa^{xy}/T$ at several $T$ for the $L=12$ cluster.
}
  \label{hdep}
 \end{center}
\end{figure}

  We show the field dependence of the longitudinal and transverse components of the thermal conductivity in Fig.~\ref{hdep}, which are even and odd functions of $\tilde{h}$, respectively.
  Figure~\ref{hdep}(a) shows $\kappa^{xx}/T$ at several $T$.
  Above $T^*$ where $\kappa^{xx}/T$ is enhanced at $\tilde{h}=0$, it decreases monotonically with increasing $\tilde{h}$.
  On the other hand, $\kappa^{xy}/T$ increases linearly to $\tilde{h}$ in the small $\tilde{h}$ region, and saturates to the quantized value $\pi/12$ for a large magnetic field, as shown in Fig.~\ref{hdep}(b).
  While decreasing $T$, the slope increases and the saturation field decreases, and finally, $\kappa^{xy}/T={\rm sgn}(\tilde{h}) \pi/12$ at $T=0$~\cite{Kitaev2006}.
  The low-field behavior indicates that $\kappa^{xy}/T \propto h^3$, where $h$ is the magnetic field in the original Hamiltonian in Eq.~(\ref{eq:H}).
  The peculiar $h$ dependence is one of the striking features in the thermal Hall effect at finite $T$.

%%%%%%%%%%%%%%%%%%%
%% discussion

  Finally, we discuss the relevance of our results to Kitaev candidate materials, such as $A_2$IrO$_3$ ($A$=Li and Na) and $\alpha$-RuCl$_3$.
  Although these materials exhibit a magnetic order below $T_{\rm N}\sim10$~K~\cite{PhysRevB.91.094422}, the Kitaev interaction has much larger energy scale compared to $T_{\rm N}$~\cite{PhysRevLett.110.097204,PhysRevB.88.035107,PhysRevLett.113.107201,1367-2630-16-1-013056,PhysRevB.91.241110,banerjee2016proximate}.
  Therefore, the present results will be compared with the experimental data in the paramagnetic state above $T_{\rm N}$ where the magnetic properties might be dominated by the Kitaev interaction.
  Our results are qualitatively consistent with the existing experimental data for $\alpha$-RuCl$_3$: $\kappa^{xx}$ has an anomalous contribution at high $T$~\cite{hirobe2016pre}, and it is suppressed by the magnetic field~\cite{leahy2016pre}.
  Furthermore, our results predict the following behaviors.
  While $T_{\rm N}$ will be suppressed by the magnetic field~\cite{yadav2016kitaev,baek2017observation}, the lower-$T$ peak of $C_v$ associated with the localized $Z_2$ fluxes, which is presumably hidden by the magnetic order in the real materials, may show up in the magnetic field.
  The absence of the peak in $\kappa^{xx}$ around this $C_v$ peak will be strong evidence of the thermal fractionalization, as the localized $Z_2$ fluxes do not carry heat.
  More prominent feature will be found in $\kappa^{xy}$; it shows a hump at a middle $T$ and a saturation to the quantized value in the low-$T$ limit.
  The $h^3$ behavior in the weak field region also awaits for experimental confirmation.

  In summary, we have investigated the thermal conductivity in the Kitaev model at finite $T$ with and without an applied magnetic field using the QMC simulations.
  We found that both longitudinal and transverse components provide good probes for fractional quasiparticles, itinerant Majorana fermions and localized $Z_2$ fluxes, inherent to the Kitaev QSL.
  The $\omega=0$ component of the longitudinal thermal conductivity exhibits a single peak structure around the higher-$T$ peak of the specific heat, which is attributed to the itinerant Majorana fermions.
  In the presence of the magnetic field, the transverse component becomes nonzero and the transverse thermal conductivity divided by $T$ exhibits a hump below the higher-$T$ peak of the specific heat, and rapidly approaches to a quantized value in the low-$T$ limit.
  We revealed that this peculiar $T$ dependence is due to thermally excited $Z_2$ fluxes and the gapped Majorana spectrum in the magnetic field.
  Moreover, we found that the transverse conductivity is induced proportional to the third order of the magnetic field, while the longitudinal one is suppressed monotonically from a nonzero value by the magnetic field.
  To our knowledge, the results provide the first quantitative theory for the thermal transport in the Kitaev model at finite $T$, which will be useful for the identification of QSL signatures in Kitaev candidate materials.

\begin{acknowledgments}
The authors thank K.~Nomura and E.-G. Moon for helpful discussions and also thank Y. Kato for a critical reading of our manuscript.
This work is supported by Grant-in-Aid for Scientific Research under Grant No. JP15K13533, JP16H00987, JP16K17747, and JP16H02206.
Parts of the numerical calculations were performed in the supercomputing systems in ISSP, the University of Tokyo.
\end{acknowledgments}

\bibliography{refs}

\end{document}